\documentclass{article}
\usepackage{spconf,amsmath,graphicx,amssymb}
\usepackage{multirow}
\usepackage[dvipsnames]{xcolor}
\usepackage{mwe,tikz}\usepackage[percent]{overpic}

\title{Using a Generative Adversarial Network for CT Normalization and its Impact on Radiomic Features}
\name{Leihao Wei$^{1,3}$ \qquad Yannan Lin$^{2,3}$ \qquad William Hsu$^{2,3,4}$ }
\address{ \small{$^{1}$ Electrical and Computer Engineering, University of California, Los Angeles} 
          \small{$^{2}$ Bioengineering, University of California, Los Angeles} \\
          \small{$^{3}$ Medical \& Imaging Informatics, University of California, Los Angeles} 
          \small{$^{4}$ Radiological Sciences, University of California, Los Angeles}
          } 

\begin{document}
\onecolumn
$\copyright$ 2020 IEEE.  Personal use of this material is permitted.  Permission from IEEE must be obtained for all other uses, in any current or future media, including reprinting/republishing this material for advertising or promotional purposes, creating new collective works, for resale or redistribution to servers or lists, or reuse of any copyrighted component of this work in other works.
\twocolumn
\maketitle
\begin{abstract}
Computer-Aided-Diagnosis (CADx) systems assist radiologists with identifying and classifying potentially malignant pulmonary nodules on chest CT scans using morphology and texture-based (radiomic) features. However, radiomic features are sensitive to differences in acquisitions due to variations in dose levels and slice thickness. This study investigates the feasibility of generating a normalized scan from heterogeneous CT scans as input. We obtained projection data from 40 low-dose chest CT scans, simulating acquisitions at 10\%, 25\% and 50\% dose and reconstructing the scans at 1.0mm and 2.0mm slice thickness. A 3D generative adversarial network (GAN) was used to simultaneously normalize reduced dose, thick slice (2.0mm) images to normal dose (100\%), thinner slice (1.0mm) images.  We evaluated the normalized image quality using peak signal-to-noise ratio (PSNR), structural similarity index (SSIM) and Learned Perceptual Image Patch Similarity (LPIPS). Our GAN improved perceptual similarity by 35\%, compared to a baseline CNN method. Our analysis also shows that the GAN-based approach led to a significantly smaller error (p-value $<$ 0.05) in nine studied radiomic features. These results indicated that GANs could be used to normalize heterogeneous CT images and reduce the variability in radiomic feature values. 
\end{abstract}
\begin{keywords}
{lung cancer, radiomics, generative adversarial networks, deep neural networks, denoising}
\end{keywords}
\section{Introduction}
\label{sec:intro}
Lung cancer is the leading cause of cancer death in the United States. The National Lung Screening Trial (NLST) clinical trial demonstrated a 20\% mortality rate reduction in patients who underwent chest low-dose computed tomography (LDCT) \cite{national2011reduced}. Radiologists routinely interpret images acquired from different scanners at varying dose levels and slice thicknesses. These differences in acquisition affect morphology and texture-based features that are used to describe pulmonary nodules, leading to inconsistencies in the detection and characterization of lesions in images. Our goal is to develop a method for normalizing heterogeneous CT scans to generate scans with common reconstruction parameters on which computer-aided diagnosis (CADx) systems can be executed. Several prior works have used deep learning to denoise LDCT images. Chen et al. \cite{chen2017low} used a residual encoder-decoder convolutional neural network to optimize mean square error (MSE) loss, while Wolterink\cite{wolterink2017generative} and Yang\cite{wgan18} used a generative adversarial network (GAN) to reconstruct normal dose CT images with better textures.
Differences in dose and slice thickness affect downstream image analysis such as radiomic feature generation and may cause CADx systems to give inconsistent results. Prior studies have shown that automated detection and nodule segmentation performance is impacted due to heterogeneous CT images\cite{wasil18,nastaran17}. Therefore, to facilitate cross-platform/protocol CT image analysis, reducing the error/variability in volume-wise CT appearance due to differences in reconstruction parameters and protocols is of special interest. One approach to addressing differences in resolution, for example, is to apply super-resolution techniques, such as asking a network to reconstruct the high-resolution image. You et al.\cite{you2019ct} showed that a CycleGAN\cite{CycleGAN2017} was capable of recovering down-sampled images. 

We introduce a novel CT image normalization method that utilizes a 3D GAN with spectral-norm. The contributions of our work are as follows: 
  
\begin{enumerate}
  \item Using a 3D GAN model, we normalize the dose and slice thickness of a CT scan simultaneously.
  \item In addition to traditional metrics such as PSNR and SSIM, we quantitatively assess perceptual quality using Learned Perceptual Image Patch Similarity (LPIPS)\cite{zhang2018unreasonable}.
  \item We show that radiomic feature variability when comparing the original heterogeneous acquisition and our normalized reconstructions have been significantly reduced. 
\end{enumerate}

\begin{figure}[!ht]
\includegraphics[width=0.48\textwidth]{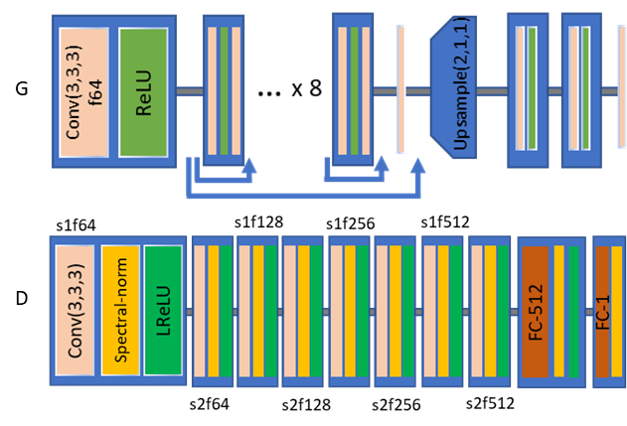}
\caption{Network structure of our GAN-based approach}
\label{fig:net}
\end{figure}

\begin{figure*}[t!]
\vspace{-5mm}
\centering
\includegraphics[width=1.24\textwidth]{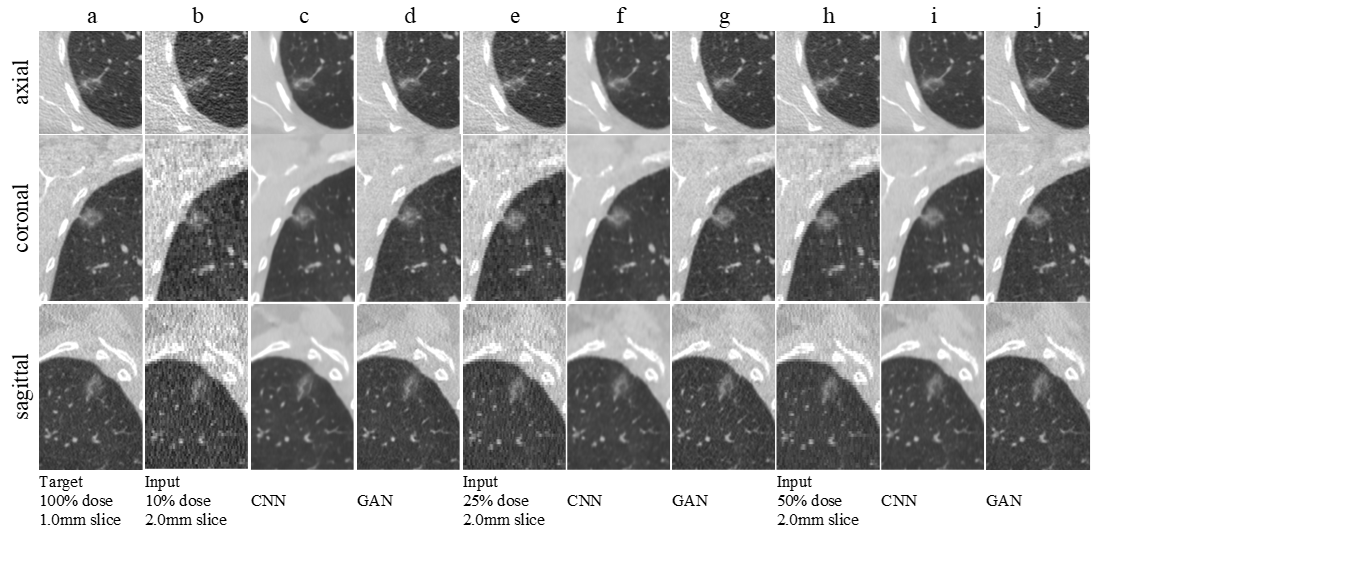}
\vspace{-12mm}
\caption{Normalization results. Each row represents the axial, coronal and sagittal view of an ROI that contains a nodule. Target images are shown in column a followed by the input images (columns b, e, h), the CNN-based (columns c, f, i) and the GAN-based (columns d, g, j) results generated from the input.}
\label{fig:norm}
\end{figure*}

\vspace{-5mm}
\section{Method}
\label{sec:method}
In prior work, GANs have been used to reconstruct photorealistic images for enhancement, denoising, and super-resolution tasks \cite{goodfellow2014generative,ledig2017photo}. Here, we developed and evaluated a 3D GAN as the basis for volumetric normalization of CT scans. Our model has a unique spectral-norm layer to further improve training stability and speed. For comparison, we also implemented a baseline model based on a convolutional neural network (CNN) with MSE loss. Each mapping between the original and normalized dose and slice thickness was trained as an independent model. We calculated image quality using peak signal-to-noise ratio (PSNR), structural similarity (SSIM) and Learned Perceptual Image Patch Similarity (LPIPS) along the axial, coronal, and sagittal views. We extracted radiomic features of a nodule in the test set and computed the absolute error between the original reduced dose/thick slice image, the CNN- and GAN-based outputs, and features generated from the normal dose/thin slice image.
\vspace{-2mm}
\subsection{Dataset}
\label{sec:dataset}
Our dataset consists of raw projection data extracted from Siemens CT scanners (Sensation 64, Definition Flash) for 40 patients who underwent chest LDCT exams. Most datasets introduce Poisson noise to the post-reconstructed image to simulate reduced doses. However, we utilize a previously validated physics-based model \cite{young2016th} that injects noise into the sinogram of each scan to simulate reduced dose levels at 10\%, 25\% and 50\% of the original LDCT acquisition. The final CT images were reconstructed via filtered back projection using a medium kernel at both 1.0mm and 2.0mm slice thicknesses with an image size of 512 $\times$ 512. We partitioned data into 30 training cases, 5 validation cases, and 5 testing cases. 
\vspace{-2mm}
\subsection{Network Architecture}
GANs consist of a generator $G$ and a discriminator $D$. The generator maps an input volume $x$ to a volume representing a standardized acquisition $G(x)$ with the reference being $y$. The discriminator $D$ is trained using both the reconstructed standardized scans and real reference scans as inputs to differentiate between $G(x)$ and $y$. $D$ constantly judges the similarity between $G(x)$ and $y$ to improve the performance of the generator. A good generative model is achieved when the discriminator can no longer distinguish between the $G(x)$ and $y$. However, GANs are notoriously difficult to train. To improve training stability, Wasserstein GAN-GP \cite{gulrajani2017improved} imposes local regularization on the discriminator to satisfy the Lipschitz continuity constraint by penalizing the gradients. Here, we use spectral-norm\cite{miyato2018spectral} to satisfy the same constraint. Spectral-norm is a robust global regularization technique that divides every weight matrix by its largest singular value, as opposed to calculating a 2nd order gradient penalty, which can be computationally expensive.  
Inspired by Enhanced Deep Residual Networks (EDSR) \cite{lim2017enhanced}, our generator contains multiple layers of residual blocks to extract features followed by an up-sampling block in the longitudinal dimension. This up-sampling block effectively increases z resolution, mitigating the partial-volume effect in thick slice scans. We use hinge loss for the discriminator to restrict $D$ to focus on hard samples that are difficult to differentiate. The generator loss function contains a $\mathcal{L}_1$ content loss and an adversarial loss. Loss functions $V_{D}(G,D)$ and $V_{G}(G,D)$ are shown in equations \ref{loss:D} and \ref{loss:G}, and training proceeds with alternating $D$ and $G$ updates, $\min_G \max_D[ V_{D}(G,D) + V_{G}(G,D)]$, where $\Theta$ and $W$ are network parameters. The network structures are summarized in figure \ref{fig:net}. For comparison, we trained the same generator network using only the $\mathcal{L}_1$ content loss as our baseline CNN model. 
\begin{equation}
\label{loss:D}
\begin{aligned}[b]
V_{D}(G,D) & = \mathop{\mathbb{E}}_{y \sim p_{y} }[\min(0,-1+D_{\Theta}(y)]  \\
           &+ \mathop{\mathbb{E}}_{x \sim q_{x} }[\min (0, -1-D_{\Theta}(G_W(x)))] 
    \end{aligned}
\end{equation}

\begin{equation}
\hspace{-2mm}
\label{loss:G}
V_{G}(G,D) = - \alpha_1 \mathop{\mathbb{E}}_{x \sim q_{x} }[D_{\Theta}(G_W(x))]  
           + \alpha_2 \mathop{\mathbb{E}}\limits_{\substack{x \sim q_{x} \\ y \sim p_{y}}} \|G_W(x) - y\|_1
\end{equation}

\vspace{-6mm}
\subsection{Model Training and Tuning}
In the training stage, we inputted randomly generated patches of size 16$\times$64$\times$64 (depth, height, width), excluding patches that were primarily outside of the body. Each voxel value was scaled from Hounsfield units to [0,1]. The outputted patch from the network has a dimension of 32$\times$64$\times$64. We trained three different GANs to accommodate the following normalization scenarios: 2.0mm slice thickness at 10\% dose, 25\% dose,  and 50\% dose to 1.0mm slice thickness at 100\% dose. The learning rate of both the generator and discriminator is set to $1e-5$. Based on hyperparameters used in \cite{miyato2018spectral}, $D/G$ update ratio was set to 1. We used Adam optimizer with $\beta_1=0.5$ and $\beta_2=0.999$. For the generator loss function weights, we used $\alpha_1=1, \alpha_2=5e-3$. These values were set using a grid search on the validation set. The batch size was set to 14, and we trained the model using 60k iterations on an NVIDIA Titan XP GPU, taking 60 hours. During inference, we used half-float precision on an input volume of 512$\times$512$\times32$ to save GPU memory. The final inference outputs are stitched back together with an overlap of 4 voxels in z direction. Processing time for the whole volume of a patient's scan with 160 slices is around 45 seconds.

\begin{table}[ht!]
\vspace{-4mm}
\centering
\caption{Image quality assessment. Axial (\textbf{Ax}), coronal (\textbf{Co}), and sagittal (\textbf{Sa}) views results are shown in each metric's sub-rows. LPIPS measures the perceptual similarity between the target and CNN-based, GAN-based normalized images (the \textbf{lower} the better). Dose-slice pairs normalization scenarios: 
\textbf{A}. 10\% dose, 2.0mm slice to 100\% dose, 1.0mm slice 
\textbf{B}. 25\% dose, 2.0mm slice to 100\% dose, 1.0mm slice
\textbf{C}. 50\% dose, 2.0mm slice to 100\% dose, 1.0mm slice}
\label{tab:metrics}
\resizebox{0.49\textwidth}{!}{
\begin{tabular}{|l|l|l|l|l|l|l|l|l|l|}
\hline
\multicolumn{2}{|l|}{\multirow{2}{*}{Metric}} & \multicolumn{2}{l|}{\textbf{A}} & \multicolumn{2}{l|}{\textbf{B}} & \multicolumn{2}{l|}{\textbf{C}} \\ \cline{3-8} 
\multicolumn{2}{|l|}{}      & CNN & GAN                 & CNN & GAN                 & CNN & GAN \\ \hline
                         
\multirow{3}{*}{PSNR(dB)}& Ax & 29.94 & 28.30             & 30.12 & 28.85             & 30.69 & 29.63 \\ \cline{2-8} 
                         & Co & 32.04 & 30.34             & 32.17 & 30.90             & 32.78 & 31.65 \\ \cline{2-8} 
                         & Sa & 30.19 & 28.55             & 30.34 & 29.08             & 30.91 & 29.84 \\ \hline
                         
\multirow{3}{*}{SSIM}    & Ax & 0.7294 & 0.6780           & 0.7380 & 0.7027           & 0.7646 &  0.7372\\ \cline{2-8} 
                         & Co & 0.7196 & 0.6655           & 0.7288 & 0.6904           & 0.7552 &  0.7249\\ \cline{2-8} 
                         & Sa & 0.7270 & 0.6762           & 0.7357 & 0.7009           & 0.7619 &  0.7342\\ \hline
                         
\multirow{3}{*}{LPIPS} 
                         & Ax & 0.3090 & 0.2147           & 0.2649 & 0.1992           & 0.2202 & 0.1844\\ \cline{2-8}  
                         & Co & 0.3286 & 0.2081           & 0.2862 & 0.1970           & 0.2514 & 0.1829 \\ \cline{2-8}
                         & Sa & 0.3059 & 0.1937           & 0.2672 & 0.1817           & 0.2342 &  0.1642\\ \hline
\end{tabular}}
\end{table}

\vspace{-8mm}
\subsection{Image Quality Assessment}
\label{methods:im}
Many metrics have been proposed to assess image reconstruction quality. Among them,  PSNR and SSIM have been widely used to measure local differences to a reference image. However, these functions are computed based on low-level features. Optimizing loss corresponding to these functions (e.g., using mean-squared error) leads to overly smoothed images with reduced texture. To better assess the image quality, we used LPIPS, a perceptual metric that utilizes a pre-trained VGG network \cite{zhang2018unreasonable} to generate similarity scores of high-level semantic features between two images. A lower LPIPS value represents a closer distance to the reference/target image. For each metric, results were calculated along the axial (x-y), coronal (x-z), and sagittal (y-z) planes. Results were averaged over all 5 scans in the test set. The same calculations were performed on all three normalization scenarios.

\vspace{-2mm}
\subsection{Radiomic Feature Error Analysis}
\label{results:radiomic}
Due to heterogeneous CT acquisitions, radiomic features extracted from images can vary widely. To evaluate the impact of our GAN-based normalization approach on radiomic feature values, we used \texttt{pyradiomics} \cite{van2017computational} to extract nine representative radiomic features. To mimic a CAD system making a diagnosis on a nodule whether being benign or malignant, we only focused on a region of interest (ROI) which occupied a patch with a dimension of 30$\times$64$\times$64. For each slice $i$ in all three normalization scenarios, we calculated a feature value $\hat{x}_i$ after normalization and the reference feature value $x_i$. The normalized error on $i$th slice was defined as $\frac{|\hat{x}_i-x_i|}{x_i}$. The same analysis was repeated for all nine features. To find the significant differences in extracted features, statistical tests were also performed as paired Wilcoxon signed-rank sum test with a significance threshold at p$<$0.05.  

\begin{figure}[!ht]
\vspace{-5mm}
  \centering   
  \begin{overpic}[width=0.48\textwidth]{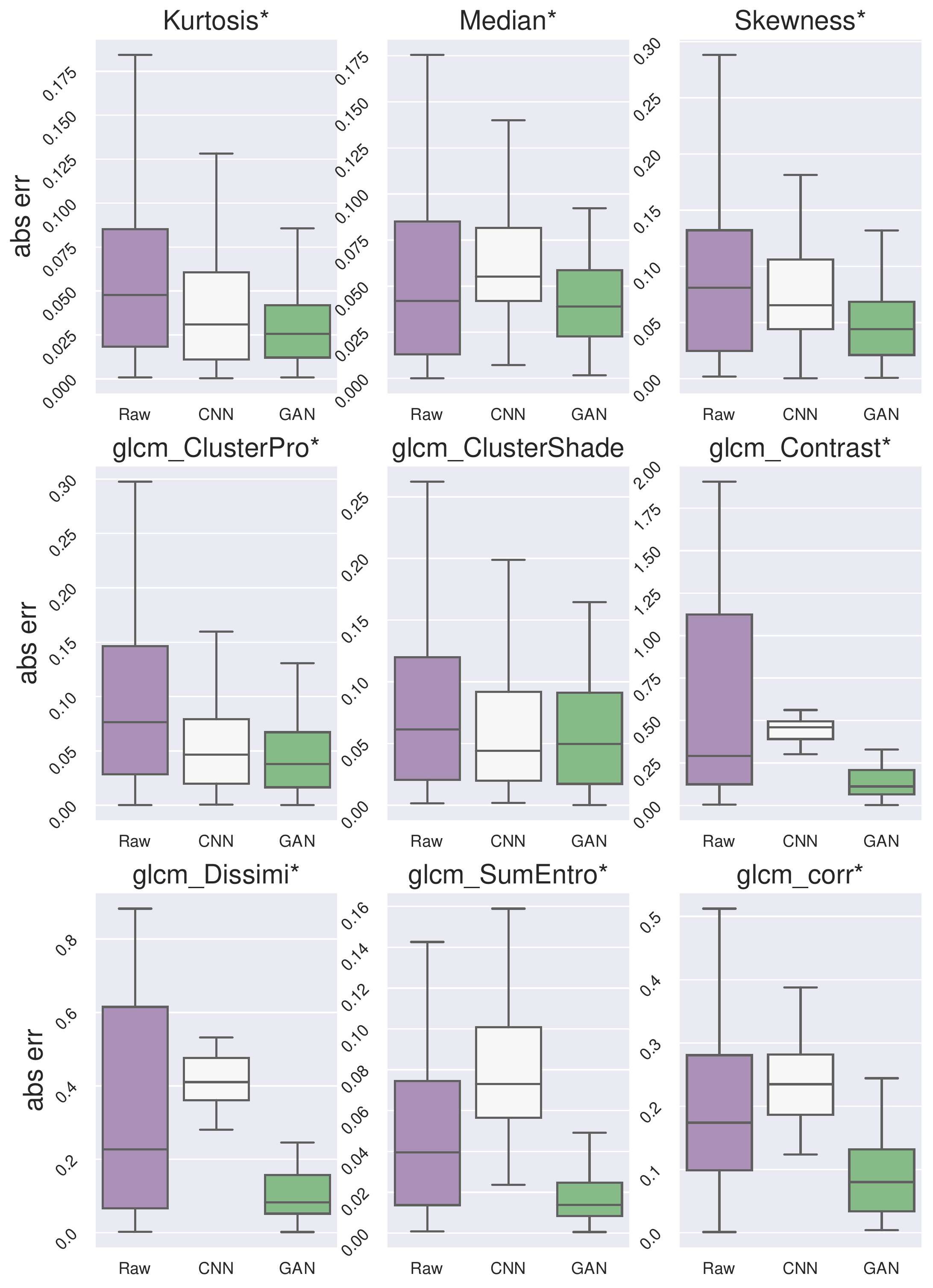}
     \put(62,88){\includegraphics[scale=0.45]{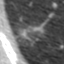}}  
  \end{overpic}
\caption{Radiomic feature errors for raw, CNN-, and GAN-based outputs across all three normalization scenarios. Radiomic features where the GAN achieved a significantly lower error compared to the CNN are denoted with an asterisk. (Wilcoxon signed rank test, p $<$ 0.05). A representative ROI from which radiomic features were calculated is shown in the top right corner.}
\label{fig:radiomics}
\end{figure}

\vspace{-2mm}
\section{Results}
\label{sec:results}
Normalization results are shown in figure \ref{fig:norm}. The ROI displayed contains a part-solid lung nodule (solid component $<$ 5mm) with poorly circumscribed margins. The displayed lung window is centered at -600 HU with a width of 1500 HU. The GAN achieved better perceptual quality than CNN with richer textures and sharp edges. Enhancement is more pronounced in the coronal and sagittal views. 
\vspace{-4mm}
\subsection{Image Quality}
Table \ref{tab:metrics} summarizes image quality results. Our GAN method achieves superior LPIPS. On average, compared to the baseline CNN, our GAN improved perceptual similarity by 35\%, 29\%, and 25\% for each of the three normalization scenarios. The largest improvement was achieved when the input was 2.0mm slice and 10\% dose (the extreme normalization scenario). However, PSNR and SSIM were lower than that of the CNN method. Although the CNN achieves higher PSNR and SSIM, the resulting images appear to lose important textures that are useful for a radiologist or CADx algorithm to inform a diagnosis. 

\subsection{Radiomics Features}
While the proposed GAN-based approach yielded superior image perceptual quality, it might also generate artifacts that were not real (e.g., unrealistic textures within the nodule). Figure \ref{fig:radiomics} shows box plots that summarize the effect of the GAN model on the error of radiomic features (e.g., comparing the value of the normalized scan to that of the standard reference scan). Among all 9 radiomic features, the GAN-based method has significantly lower mean feature errors compared to the non-normalized, where the largest p-value is 0.035. The GAN-based method also has a lower error compared to CNN-generated scans in 8 features marked with an asterisk.

\vspace{-4mm}
\section{Discussion}
By normalizing heterogeneous CT scans to a standardized dose and slice thickness, we show that the GAN-based method not only leads to a better perceptual appearance in the normalized images but also results in reduced variability in radiomics features. These results are potentially valuable for developing CADx systems that can achieve more consistent detection and classification performance when presented with images acquired using different CT hardware platforms/protocols. In addition, different from prior work which only focused on denoising or super-resolution independently, our method performs these tasks simultaneously.

Some limitations of this work include:
1. We have not evaluated how normalization would impact nodule detection and classification tasks by CADx. Since the ultimate goal is to improve the CADx system performance by normalizing scans from different protocols or reconstruction parameters, we need to further evaluate the impact of the GAN-based approach to these specific diagnosis tasks. We intend to assess the impact of our normalization approach on the performance of existing systems, including our own \cite{yannan}. 2. Since reconstruction kernel also plays an important role in image textures and thus impacts CADx performance on texture-based feature extraction, we also plan to investigate the feasibility of normalizing different kernels. 3. We have not yet explored how to tune the GAN to faithfully generate features within and immediately surrounding the nodule, which are the critical areas to enhance for the target classification task.

\vspace{-4mm}
\section{Acknowledgements}
The authors acknowledge funding support for this work from the National Cancer Institute of the National Institutes of Health under award R01 CA210360 and the National Science Foundation under grant \#1722516. We thank Drs. Michael McNitt-Gray and John Hoffman for providing access to their physics-based simulated chest CT dataset. Computing resources were provided by a donation of a Titan Xp GPU by NVIDIA Corporation. The content is solely the responsibility of the authors and does not necessarily represent the official views of sponsor agencies.

\bibliographystyle{IEEEbib}
\bibliography{root}

\end{document}